# Synthesis of Luminescent Eu defects in diamond

Andrew Magyar[1], Wenhao Hu[2], Toby Shanley[3], Michael E. Flatté[2], Evelyn Hu[4], Igor Aharonovich[3*]

1. Center for Nanoscale Systems, Harvard University, Cambridge MA 02138 USA
2. Dept. of Physics and Astronomy, University of Iowa, Iowa City, IA 52242, USA
3. School of Physics and Advanced Materials, University of Technology Sydney, Ultimo, NSW 2007, Australia
4. School of Engineering and Applied Sciences, Harvard University, Cambridge MA, 02138 USA

Corresponding author: * igor.aharonovich@uts.edu.au

**Abstract:**

Lanthanides are vital components in lighting, imaging technologies and future quantum memory applications due to their narrow optical transitions and long spin coherence times. Recently, diamond has become a preeminent platform for realization of many experiments in quantum information science. In this work, we demonstrate a promising approach to incorporate Eu ions into single crystal diamond and nanodiamonds, providing a means to harness the exceptional characteristics of both lanthanides and diamond in a single material. Polyelectrolytes are used to electrostatically assemble Eu(III) chelate molecules on diamond and subsequently chemical vapor deposition is employed for the growth of a high quality diamond crystal. Photoluminescence, cathodoluminescence and time resolved fluorescence measurements show that the Eu atoms retain the characteristic optical signature of Eu(III) upon incorporation into the diamond lattice. Computational modelling supports the experimental findings, corroborating that $Eu^{3+}$ in diamond is a stable configuration within the diamond bandgap. The versatility of the synthetic technique is further illustrated through the creation of the well-studied Cr defect center. Together these defect centers demonstrate the outstanding chemical control over the incorporation of impurities into diamond enabled by the electrostatic assembly together with chemical vapour deposition growth.

**R**are earth ions are vital components of many modern technologies including lasers, light emitting diodes and communications[1-5]. The storage of quantum bits is critical to future quantum information technologies. Lanthanides exhibit remarkably long nuclear spin coherence times; europium exhibits nuclear spin coherence times that are extremely long (T2 ~ minutes T1 ~ weeks), making it a superb candidate for such quantum memories or for applications in magnetic sensing[6-8]. In addition, Eu(III) has drawn attention for its photostable, narrowband fluorescence lines across in the visible. Coherent manipulation of several spin selective transitions allows optical detection of the nuclear spin state enabling their implementation as local NMR probes[9-11].

Currently, diamond is one of the preeminent platforms for high resolution magnetometry and quantum information processing, owing to the nitrogen vacancy (NV) defect[12-15]. Diamond possesses a very large bandgap (5.5 eV) that can accommodate a variety of optically active defects. Furthermore, the progress in growing isotopically pure diamond, (by minimizing $^{13}$C isotopes) has recently enabled very long spin coherence times of the NV[16, 17]. The chemical stability of diamond means that nanodiamonds exhibit excellent photostability and cytotoxicity – making them promising as bio-labels and biosensors. Combining the extreme properties of diamond with the unique atomic and optical properties of rare earth ions is an important first step towards an all diamond solid state platform for quantum information processing and sensing.

In this work we demonstrate computationally that Eu is stable within the diamond matrix and show that the native f-orbital transitions of a Eu(III) atom should be preserved. We then present a robust method to incorporate Eu ions into diamond using chemical self assembly and growth. A layer-by-layer approach is employed to first electrostatically assemble a positively charged polymer on an oxidized diamond surface and then assemble a negatively charged Eu complex on top of the polymer. The coated diamond is then transferred into a chemical vapor deposition chamber for a subsequent overgrowth of diamond to generated a Eu-doped layer.

This technique has the advantage that nearly any atomic species that can be electrostatically assembled on the diamond surface can be incorporated in diamond. Currently ion implantation is commonly used to dope solids with external ions, however this approach can irreversibly damage the host matrix. In technique presented in this work, the doping process occurs using a bottom up method that does not damage the host matrix and preserves the atomic transitions of the alien atom.

**Results**

**Simulations.** To understand the feasibility of incorporating lanthanide defects into the diamond lattice, the energetics of a Eu defect within the diamond lattice was simulated. The formation energy is calculated for Eu in diamond using density functional theory within the local spin density approximation as implemented in the WIEN2K code[18]. A 64-site supercell is considered, corresponding to a cube of 2×2×2 nonprimitive cubic unit cells of diamond, with a europium atom substituted in the center and next to 1, 2 or 3 carbon vacancies. There is insufficient space in the crystal lattice for Eu to substitute for C without any vacancies. A k-point mesh of 4×4×4 is used to sample the irreducible Brillouin zone. The Coulomb interaction parameter U and exchange parameter J are set to 7.397 eV and 1.109 eV respectively, and the spin-orbit interaction is neglected[19]. The formation energy and 4f-shell occupation for charge states of Eu ranging from -2 to +4 have been calculated for a Fermi energy at the top of the valence band, and are shown in Table 1.

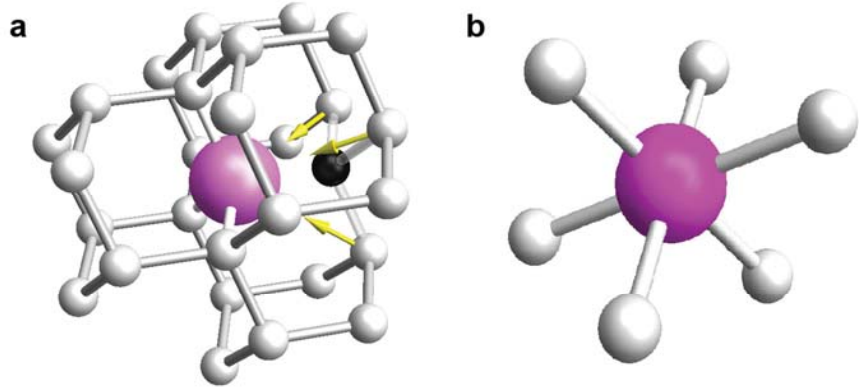

**Figure 1. The relaxation process in one vacancy structure. (a)** After the substitution of Eu, three nearest neighbor carbons of europium and another three carbons bonded with the taken carbon will relax towards the Eu. **(b)** A defect center with a six-fold coordinate shell of carbon atoms will be formed after relaxation.

| Charge State | 1 vacancy | | 2 vacancies | | 3 vacancies | |
|---|---|---|---|---|---|---|
| | $E_f$ (eV) | 4f Occ | $E_f$ (eV) | 4f Occ | $E_f$ (eV) | 4f Occ |
| -2 | 12.25 | 5.8736 | 13.38 | 6.3537 | 13.67 | 5.8753 |
| -1 | 11.93 | 6.3709 | 11.9 | 5.9843 | 13.32 | 6.4345 |
| 0 | 11.58 | 6.0934 | 12.04 | 6.0344 | 13.37 | 6.1197 |
| 1 | 11.53 | 6.3377 | 12.14 | 6.1259 | 13.25 | 6.4634 |
| 2 | 12.16 | 6.5032 | 13.47 | 6.0614 | 14.57 | 6.6814 |
| 3 | 13.77 | 6.3137 | 15.66 | 6.0855 | 16.91 | 6.6093 |

**Table 1: Formation energy and 4f electron occupation of the Eu defect**

It is known that correlation effects in the 4f series stabilize the f-shell occupation relative to 5s and 5p occupation[20], even as the charge state of the ion varies and for 4f levels in the band gap. Hence we find that the average 4f occupation remains relatively stable (near that expected for the +3 state of the isolated atom) even as the actual defect charge state varies from -2 to +4. A single vacancy is found to be the most stable configuration with Eu; in the relaxed positions of the atoms the carbon neighbors of the vacancy relax towards the Eu, forming a six-fold coordinated shell of carbon atoms surrounding the Eu, as shown in Figure 1.

**Results**

**Experimental synthesis.** In preliminary studies we were unable to create luminescent europium defects via ion implantation of Eu into the diamond crystal. To eliminate the implantation-related damage, a technique was developed to introduce Eu into the diamond matrix during microwave CVD growth. The process used to prepare a diamond having a monolayer of Eu for growth is shown in Figure 2.

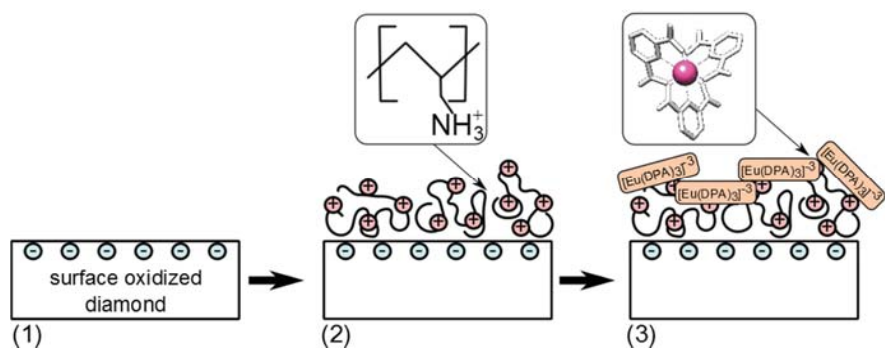

**Figure 2. An overview of the process used to assemble the Eu precursor on a bulk diamond prior to growth.** (1) The surface of the diamond is oxidized using a boiling tri-acid etch, (2) A positively charged polymer, PAH, is assembled electrostatically on the negatively charged diamond (3) The Eu precursor, [Eu(III) tridipicolinate]$^{-3}$ is assembled on the positively charged polymer.

A monolayer was desired to minimize the impact of the Eu on the growth conditions. To assemble this monolayer, first the diamond was surface oxidized using a boiling tri-acid etch (1:1:1 nitric:sulfuric:perchloric acids) to create a negatively charged surface. Next, polyallylamine hydrochloride, a positively charged polymer, was assembled on the negatively charged diamond surface. Finally a negatively charge europium complex, Eu(III) tri-(2,6-pyridine dicarboxylic acid) (EuDPA), was assembled on the positively charged polymer layer.

The use of a chelated Eu complex and the polymer layer may seem superfluous, however they are both critical to the assembly of an Eu(III) monolayer on the diamond. The surface of oxidized diamond is only significantly negatively charged at above pH 5. Unfortunately, at elevated pH Eu(III) rapidly forms oxides and hydroxides and precipitates out of solution. Eu oxides are stable even at the elevated temperatures present during CVD diamond growth, making them non-ideal for the facile incorporation of Eu into diamond.

Three different techniques were used to prepare Eu:nanodiamond substrates for growth: (1) 5 nm nanodiamond seeds were mixed with an aqueous solution of EuDPA and deposited on $SiO_2$ on Si; (2) nanodiamonds were oxidized in boiling acid, coated with PAH and then coated with EuDPA and then deposited on $SiO_2$ on Si; (3) the $SiO_2$ substrate was treated with hot piranha to create a highly negatively charge surface, then PAH was assembled on the substrate followed by the assembly of EuDPA and lastly nanodiamonds were dispersed on the coated substrate.

The diamond growth was carried out in a conventional microwave assisted chemical vapor deposition reactor under pressure of 60 torr and 900 W microwave plasma. The duration of the growth was approximately 10 minutes to grow individual nanodiamonds. Post-growth the diamond specimens were characterized by photoluminesence spectroscopy. Samples were excited with a 395 nm pulsed laser. The fluorescence for a bulk diamond sample containing Eu is shown in Figure 3a. The fluorescence of the Eu in diamond (3a, bottom, blue curve) differs from that of the precursor (3a, top, red curve). The strongest peak, corresponding to the $^5D_0 \rightarrow {^7F_2}$ transition, is a singlet in the precursor and a multiplet in the diamond sample. The $^5D_0 \rightarrow {^7F_1}$ and $^5D_0 \rightarrow {^7F_4}$ transitions are significantly broadened, perhaps also as a result of peak splitting. Lastly, the Eu in diamond exhibits an additional singlet at 582 nm that is not present in the precursor, likely corresponding to the $^5D_0 \rightarrow {^7F_0}$ transition.

Photoluminescence measurements of nanodiamonds grown via each technique showed that all three techniques lead to similar Eu defects in the nanodiamond. However, the third technique proved most effective. Simply mixing the Eu complex with the nanodiamonds (technique 1) led to the formation of large unknown crystallites that produce a broad background fluorescence signal while the Eu-associated fluorescence signal was significantly weaker for the nanodiamonds that were coated with Eu (technique 2), perhaps because of less total Eu available during growth.

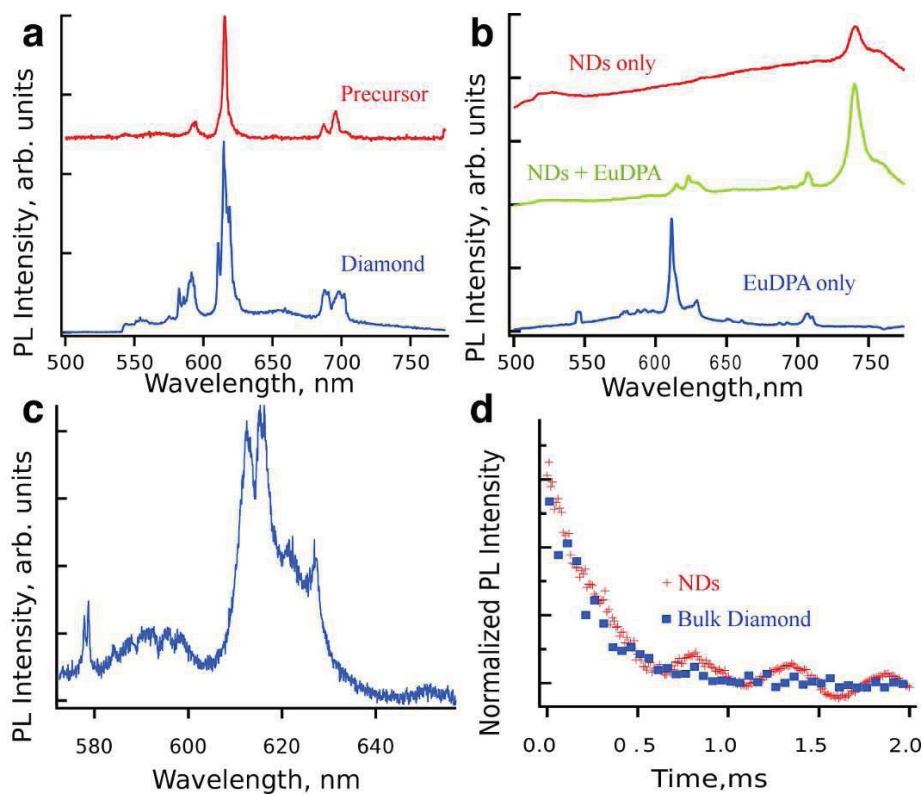

**Figure 3: Photoluminescence of europium defects (a)** PL from the EuDPA precursor (top) and Eu defects in bulk diamond after growth (bottom). **(b)** PL of nanodiamonds grown without Eu (top), nanodiamonds grown with EuDPA (middle) and EuDPA grown without nanodiamonds (bottom). **(c)** A high resolution PL spectrum of the Eu defect in nanodiamonds revealing the line splitting of the $^5D_0 \rightarrow {}^7F_1$ transition. **(d)** Fluorescence decay of Eu defects in bulk diamond (blue squares) and nanodiamonds (red cross).

A fluorescence spectrum of Eu nanodiamonds is shown in Figure 3b (middle). Spectra of nanodiamonds grown without the Eu precursor (3b top) and a sample with the Eu precursor but no nanodiamond seeds (3b bottom) are shown for comparison. The spectrum for the Eu nanodiamonds is distinctly different for that of the precursor grown without any seeds. The peak at 738 nm seen in the samples grown with a nanodiamond seed (3b top and middle) is characteristic of silicon vacancies (SiV) in diamond, likely introduced due to the Si substrate on which the nanodiamonds were grown[21]. A higher resolution spectrum of the $^5D_0 \rightarrow {}^7F_2$ transition for the nanodiamond sample is shown in Figure 3c, detailing the splitting of this peak.

Fluorescence lifetime measurements, shown in Figure 3d were recorded from both the Eu bulk and Eu nanodiamond samples. The Eu defect in the nanodiamond sample exhibited a lifetime of 325 μs while the lifetime of the defects in the bulk sample were slightly longer,

391 µs. The shorter lifetime in the nanodiamonds is likely to arise from various non radiative decay channels present in the crystal. Remarkably, however, the measured lifetime of the Eu defects in diamond is much shorter than other reported Eu related complexes[8, 22]. This is attributed to the wide band gap of the diamond lattice, which localizes the wave functions of the 4f levels strongly and thus enhances the emission.

The nanodiamonds are smaller than the diffraction limit of light, so to further verify that the Eu(III) fluorescence was indeed associated with individual nanodiamonds, cathodoluminescence (CL) studies of single nanodiamonds using 10 keV electron beam were carried out *insitu* within a scanning electron microscope (SEM). The light is collected using a parabolic mirror and directed into a spectrometer. Figure 4a shows a SEM image of several individual nanodiamonds that were grown on top of the Eu containing polymer. Figure 4b shows the recorded CL signal unveiling the common A band at 460 nm (attributed to nitrogen aggregates in diamond), the narrow peak at 738 nm[23], attributed to the SiV and the emission centered at 612 nm, associated with a Eu complex in diamond. Interesting to note, that even in CL measurements the Eu signal in diamond is replicated, indicating the possible transition pathways from the excited state to the Eu levels. Due the complex electron beam dynamics, not all color centers are visible in CL. The availability of imaging Eu defects with CL opens important possibilities for detailed characterization of these defects using time resolved high resolution CL methods.

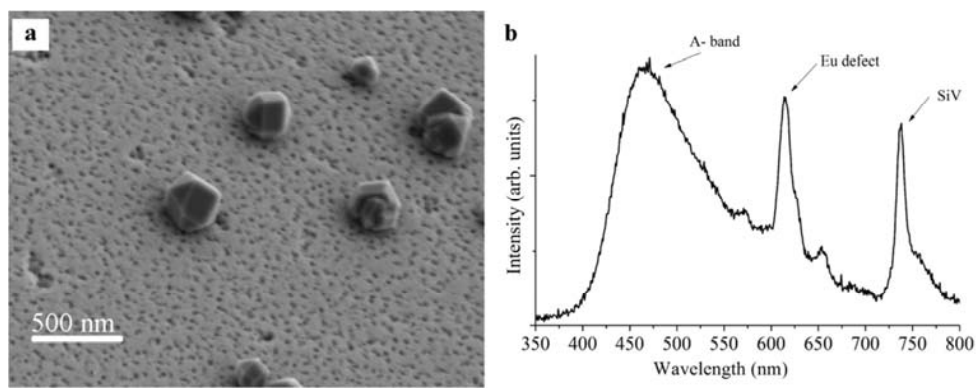

**Figure 4. Cathodoluminescence imaging of Eu defects in nanodiamonds. (a)** SEM image of the nanodiamonds containing Eu defects. **(b)** CL signal recorded from an individual nanodiamonds, resolving the A band fluorescence (broad emission at ~ 460 nm), the SiV defect (~ 738 nm) and the Eu line (~612 nm).

**Discussion**

To demonstrate the broad utility of this technique for creating color centers in diamond, chromate was electrostatically assembled on the surface of a bulk diamond, as described for the EuDPA complex above. Chromate was selected because chromium centers are a well-studied diamond defect that exhibits fluorescence that is distinct from the native ion. After growth, photoluminescence measurements, revealed a very sharp resonance centered at 761.14 nm, as shown in Figure 5, in accordance to the previous results on chromium related color centers in diamond[24, 25]. Therefore, the chemical technique is a viable rote for engineering color centers in diamond and potentially other wide band gap semiconductors.

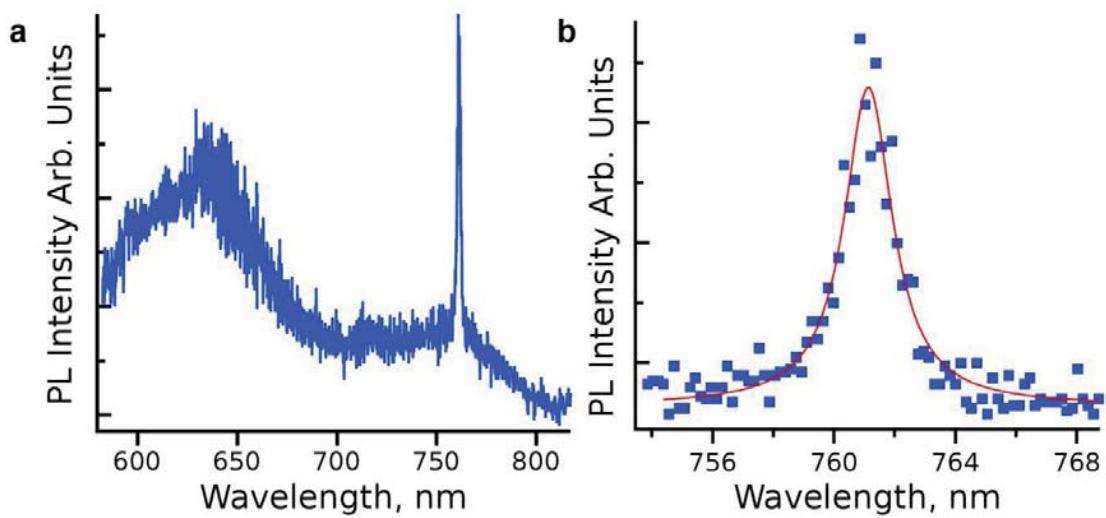

**Figure 5: Photoluminescence from chromium defects in diamond, created by electrostatically assembling chromate on a bulk diamond. (a)** Room temperature PL spectrum showing a bright narrowband line at 760 nm, corresponding to a Cr defect. **(b)** A magnification of the spectral region around 760 nm, showing the Cr-defect resonance.

To summarize, we presented a chemical method to engineer Eu related defects in diamond. The incorporation of Eu into diamond while retaining the optical transitions associated with Eu(III) is an important first step towards developing an all diamond platform for quantum memory[26] or computation. Since the process involves the introduction of the Eu during growth there is no residual damage to the diamond lattice as is caused via ion implantation. Moreover, as demonstrated with Cr, this versatile method can be extended to many other atomic species. Further exploration of other lanthanides such as Pr and Gd may be of particular interest given the unique optical and magnetic properties of these species.

**Methods**

**Preparation of Eu(III) tri(2,6-pyridine dicarboxylic acid)**
Dipicolinic acid (DPA) (0.5 g, Sigma Aldrich) was dissolved in 10 mL of water. NaOH (3.3 mL, 2M) was added to DPA solution deprotonate the acid. The solution was brought to a boil while stirring magnetically until the DPA was dissolved. A solution of $EuCl_3 \cdot 6H_2O$ (0.267 g in 1 mL) was added to the DPA solution. The solution was allowed to cool and 0.2 M NaOH was added dropwise until the solution reached pH 8. Addition of NaOH too rapidly will lead to the irreversible formation of a white precipitate of europium oxide or hydroxide. The solution was left in a fume hood to crystallize and yellow EuDPA crystals were removed from the mother liquor, rinsed and dried.

**Deposition of EuDPA on bulk diamond.**
A 3 x 3 mm diamond (Element Six) was cleaned in boiling tri-acid (1:1:1 $HNO_3:HClO_4:H_2SO_4$, 133 μL each). The solution was heated in a sand bath to ~300°C in small round bottom flask with condenser. The diamond was left to sit for 1 hour and the heat was turned off. The diamond was allowed to cool overnight in the acid solution and rinsed 3x in ultrapure water.

The diamond was incubated overnight in a solution of polyallylamine hydrochloride (PAH) (50k MW, 0.1 M by monomer, Sigma Aldrich) adjusted to pH 8.0. A pH of 8.0 or higher is critical to having a negatively charged diamond surface. Excess PAH was rinsed off with ultrapure water. To coat with EuDPA, the PAH treated diamond was incubated for three hours with a 0.1 M solution of EuDPA adjusted to pH 3.8. The diamond was incubated for 3 hours and rinsed thoroughly.

**Co-deposition of EuDPA and nanodiamonds**
A dilute solution (1 mg/ml) of seed nanodiamonds (1 mL, 4-6 nm) was centrifuged (10000 RPM, 15 minutes) to form a pellet. The nanodiamond pellet was rinsed 3x with ultrapure water and then resuspended in about 1 mL of the same. A 50 µL aliquot of the solution of cleaned NDs was co deposited with 50 µL of EuDPA solution (10 mg/mL, no pH adjustment) on a cleaned 1 cm square piece of a polished Si wafer. After growth under conditions described in the main manuscript the unknown crystallites were observed together with the nanodiamonds.

**Coating NDs with EuDPA**
Nanodiamonds (500 – 1000 nm) were oxidized in boiling triacid (as described for bulk diamond) for 6 hours and allowed to cool fully. The solution of NDs and triacid was carefully centrifuged and the acid solution removed. The NDs were then rinsed thoroughly with ultrapure water to remove residual acid. The NDs were then resuspended at a concentration of 10 mg/mL in an aqueous solution of PAH (50k MW, 0.1 M by monomer, adjusted to pH 8.0). The PAH coated NDs were centrifuged to form a pellet, rinsed with ultrapure water and repelleted. The pellet of PAH coated NDs was resuspended in 10 mg/mL EuDPA (1 mL) and incubated overnight. The coated NDs were again rinsed with ultrapure water and deposited on a piece of Si for growth.

**Depositing Eu on $SiO_2$**
A 1 cm square piece of thermally grown $SiO_2$ on Si was oxidized in a hot piranha bath. The substrate was incubated with an aqueous solution of PAH (0.1 M by monomer, adjusted to pH 8) for 1 hour. The substrate was rinsed with copious ultrapure water to remove unbound polymer and then incubated with a solution of EuDPA (10 mg/mL, pH 3.9) for 1 hour. The sample was again rinsed with ultrapure water to remove excess EuDPA. Prior to growth 4-6 nm nanodiamond seeds were deposited on the substrate.

**Coating bulk diamond with chromate**
A 3 x 3 mm bulk diamond was oxidized via triacid etch (described above). The specimen was rocked in 1 mL of PAH solution (0.2 M, adjusted to pH 8.0 w/ 1M NaOH) overnight. The specimen was rinsed thoroughly with ultrapure water. The PAH coated diamond was incubated with 1 mL of aqueous solution of potassium chromate (0.1M, adjuted to pH 5) for 3 hours and then rinsed thoroughly with ultrapure water.

**Diamond Growth**

The diamond growth was carried out in a conventional microwave assisted chemical vapor deposition reactor (Seki Technotron, Model: AX5010-INT) under pressure of 60 torr and 900 W microwave plasma. The methane to hydrogen flow ratio was 4:400 standard cubic centimeter per minute. No heating stage was used during the growth.

**Optical measurements**

The optical measurements were performed at room temperature using a frequency doubled titanium sapphire laser (Mira 900), emitting at 395 nm. The excitation was passed through a dichroic mirror onto a high (0.9) numerical aperture objective (Nikon). The light was collected through the same objective and directed onto a spectrometer. Confocal scanning was achieved using a 2D scanning piezo stage (Thorlabs).

**Author Contributions**

AM conceived the idea. AM and IA performed the sample preparation and characterization. MF and WH performed the computational analysis. TS performed the CL measurements. The manuscript was written by IA and AM with contributions of all authors. All authors discussed the results and analysed the data. All authors have given approval to the final version of the manuscript.

**Acknowledgments**

We thank Jonathan Lee, Thomas Babinec and Huiliang Zhang for useful discussions. We thank Abel and Albert for substrate preparation. We thank Brett Johnson and Jeffrey McCallum for their assistance with ion implantation and The Department of Electronic Materials Engineering at the Australian National University for the access to the implantation facilities. Dr. Aharonovich is the recipient of an Australian Research Council Discovery Early Career Research ward (Project No. DE130100592). WH, MF and EH acknowledge the support of an AFOSR MURI. Part of this work was performed at the Center for Nanoscale Systems (CNS), a member of the National Nanotechnology Infrastructure Network (NNIN), which is supported by the National Science Foundation under award no. ECS-0335765. CNS is part of Harvard University.